\documentclass[sigconf]{acmart}

\acmConference[Conference acronym 'XX]{Make sure to enter the correct
  conference title from your rights confirmation emai}{June 03--05,
  2018}{Woodstock, NY}
\AtBeginDocument{%
  \providecommand\BibTeX{{%
    \normalfont B\kern-0.5em{\scshape i\kern-0.25em b}\kern-0.8em\TeX}}}

\setcopyright{none}
\usepackage{shortcuts}
\usepackage{multirow}
\usepackage{pifont}
\usepackage{graphicx}
\usepackage{subcaption}
\usepackage[strict]{changepage}
\usepackage{tikz}

\usepackage{framed}
\definecolor{formalshade}{RGB}{248,248,255}
\definecolor{LightSkyBlue}{RGB}{135,206,250}

\newenvironment{formal}{%
  \MakeFramed{\advance\hsize-\width\FrameRestore}%
  \noindent\hspace{-4.55pt}
  \begin{adjustwidth}{}{7pt}%
}
{%
  \end{adjustwidth}\endMakeFramed%
}

\let\emptyset\varnothing

\usepackage{makecell}

\usepackage{enumitem}	
\setlist{leftmargin=1.75em}
\begin{document}

\title{Exploring \gpt App Ecosystem: Distribution, Deployment and Security}

\author{Chuan Yan}
\orcid{0000-0003-4855-1912}
\affiliation{%
  \institution{University of Queensland}
  \country{Australia}
}

\author{Ruomai Ren}
\affiliation{%
   \institution{University of Queensland}
   \country{Australia}
}

\author{Mark Huasong Meng}
\affiliation{%
   \institution{Technical University of Munich}
   \country{Germany}
}

\author{Liuhuo Wan}
\affiliation{%
   \institution{University of Queensland}
   \country{Australia}
}

\author{Tian Yang Ooi}
\affiliation{%
   \institution{University of Queensland}
   \country{Australia}
}

\author{Guangdong Bai}

\affiliation{%
  \institution{University of Queensland}
  \country{Australia}
}
\authornote{Corresponding author.}

\renewcommand{\shortauthors}{Yan et al.}

\begin{abstract}

\gpt has enabled third-party developers to create  plugins to expand \gpt's capabilities. 
These plugins are distributed through OpenAI's plugin store, making them easily accessible to users. 
With \gpt as the backbone, this app ecosystem has illustrated great business potential by offering users personalized services in a conversational manner.
Nonetheless, many crucial aspects regarding app development, deployment, and security of this ecosystem have yet to be thoroughly studied in the research community, potentially hindering a broader adoption by both developers and users.
In this work, we conduct the first comprehensive study of the \gpt app ecosystem, aiming to illuminate  its landscape for our research community. 
Our study examines the distribution and deployment models in the integration of LLMs and third-party apps, and assesses their security and privacy implications.
We uncover an uneven distribution of functionality among \gpt plugins, highlighting prevalent and emerging topics. 
We also identify severe flaws in the authentication and user data protection for third-party app APIs integrated within LLMs, revealing a concerning \emph{status quo} of security and privacy in this app ecosystem. 
Our work provides insights for the secure and sustainable development of this rapidly evolving ecosystem.

\end{abstract}

\maketitle
\section{Introduction}
\gpt is a flagship large language model (LLM) product of OpenAI~\cite{openai2023creating} launched in 2023. \footnotetext{This paper has been accepted by the 39th IEEE/ACM International Conference on Automated Software Engineering (ASE 2024).}
It represents the state-of-the-art advancement in AI-driven natural language processing (NLP) technology. 
\gpt has exhibited proficiency in language comprehension and text generation by leveraging the transformer neural network architecture~\cite{vaswani2017attention} and training based on extensive corpora encompassing public knowledge and real-world dialogues on the Internet.
It can closely mimic genuine human conversions through context-aware interactions, demonstrating an extraordinary potential to provide a one-stop solution for personal AI assistants.
As of June 2024, \gpt has managed to attract an ever-growing user base reaching over 180 million users~\cite{duarte2024number}, which is achieved in less than 20 months since its debut to the public. 

Inspired by the tremendous success of the Android app ecosystem, \oai
launched the plugin store~\cite{shaikh2023chatgpt} in March 2023 to enhance the \gpt's capabilities and offer more customized experiences. 
It was made available to a selected group of users, and then to all \gpt Plus subscribers in May 2023~\cite{harshini2023openai}, 
marking a significant step in developing an LLM app ecosystem. 
These plugins enable \gpt to link the current conversation to external data sources and services such as a mobile app with internet access. 
Their functionality spans a wide range of applications, including querying real-time data such as weather conditions, airfare and hotel rates, as well as providing domain-specific professional services such as crafting magical stories. 
This highlights \oai's ambition to offer highly personalized AI services and establish \gpt as the backbone of an open app ecosystem.

Unlike mobile app development that involves full-stack implementation, \oai intends to build a low-code development paradigm.
With the powerful \gpt serving all user interactions, developers only need to upload their APIs according to the provided specifications and configure the corresponding manifest files. 
For example, when a user asks, \emph{``What is the weather like in New York on June 7th?''}, \gpt automatically collaborates with the relevant weather plugin to understand the user's prompt and constructs API requests to the plugin. 
The constructed requests are then sent to the plugin server, and based on the response from the server, \gpt produces a natural language response for the user.
This simple and lightweight development paradigm has received an immediate welcome from developers, illustrating great business potential.
Nonetheless, this app ecosystem is still in its nascent stage, although it has undergone several waves of evolution since the debut of the plugin store.
Many crucial aspects regarding \emph{characteristics of existing apps, app development, deployment and distribution mechanisms}, and \emph{security and privacy implications} have yet to be thoroughly studied in the research community.

\paragraph{Our work} 
To address this gap, we conduct a comprehensive study of the \gpt app ecosystem. 
Our study focuses on three key research questions (RQs) that are of major concern to app users, developers, and store operators, including \emph{what are the characteristics of the plugins available in the store (\textbf{RQ1})}, \emph{what are the deployment model and runtime execution model that integrate third-party apps and LLMs (\textbf{RQ2})}, and \emph{what are the security and privacy issues associated with the integration (\textbf{RQ3})}. 

\emph{\textbf{RQ1. Characterizing existing \gpt plugins}}. We collect all currently available plugins from the store (overall \allplugin), and understand their functionality based on the descriptive texts released in the store. 
The challenge to overcome in this process includes the variety of plugins and the lack of well-labeled data. 
To address this, we employ a zero-shot classification~\cite{xian2016latent} which can accurately categorize new, unseen data without the need for additional training.
Our study provides app users with a comprehensive overview of service types that are accessible to them. 
It also reveals an uneven functionality distribution, with more than half of the plugins concentrated in five categories of data \& research, tools, developer \& code, business, and entertainment, providing a useful guideline for plugin developers and store operators. 
This component is detailed in \textbf{Section~\ref{sec:RQ1}}. 

\emph{\textbf{RQ2. Understanding plugin deployment and execution models}}. 
We reverse engineer the plugin deployment and execution mechanisms by analyzing runtime traffic and data flow. 
One significant challenge in this process is that LLMs are essentially black boxes, which makes it difficult to accurately capture and interpret the runtime workflow and data flow. 
To tackle this challenge, we create our own apps and conduct testing around them. 
Through this, we identify the resources and sensitive data involved in plugin deployment and execution, such as user prompts, API keys, and access tokens. 
Our analysis provides an in-depth understanding of the internals of plugins' deployment and execution, 
serving as the foundation for our security assessment and future research in this area. 
We detail our analysis for RQ2 in \textbf{Section~\ref{sec:RQ2}}. 

\emph{\textbf{RQ3. Assessing security and privacy implications}}. 
Based on the plugin deployment and execution mechanisms, we design a three-layer security assessment model. 
This model addresses the perspectives of both developers and users, by examining potential exposure points of deployment platform resources and user data involved in the plugin execution workflow.  
Our assessment reveals three potential security issues associated with the plugin deployment and execution, namely \emph{credential leakage}, \emph{data provision inconsistency}, and \emph{broken API authorization}. 
Among \allplugin plugins, 173 plugins have broken access control~(BAC) vulnerabilities, 69 exhibit inconsistencies in data provided to users and \gpt, and 368 are subject to leaking developer credentials such as API keys, API locations, and OAuth tokens. 
Regarding user data protection, we examine the legal documents related to user data collection as mandated by \gpt store. 
We find that 271 plugins provide legal document links that are inaccessible. 
These findings reveal a worrisome prevalence of security and privacy flaws among ChatGPT plugins. 
This analysis and the responsible disclosure of our findings are detailed in \textbf{Section~\ref{sec:exposure}}.

\paragraph{Contributions}
The main contributions of this work are as follows.
\begin{itemize}
    \item \textbf{A comprehensive characterization of \gpt plugins.}
    To the best of our knowledge, we are the first to systematically study the existing apps in the \gpt plugin store. 
    We summarize functionalities provided by these apps, offering an overview to app users. 
    We also highlight the uneven distribution of plugin categories with prevalent and emerging topics, which can serve as a reference for developers in deciding their future endeavors, and for store operators in personalizing services such as app recommendation. 
    \item \textbf{A systematic security assessment and practical impact.}
    We reveal the deployment and runtime execution mechanisms of \gpt plugins for the first time. 
    Based on that, we propose a three-layer security assessment to evaluate the resource and data exposure associated with \gpt plugins. 
    Our approach can effectively detect five predefined types of exposures that may exist in plugins, leading to numerous vulnerable plugins. 
    We also discover that many plugins fail to adhere to \oai's regulations regarding providing legal documents.
    Our findings are responsibly disclosed to \gpt, who has put efforts into the enhancement. 
    
    \item \textbf{Revealing the \emph{status quo} and development trajectory of \gpt plugin store.} 
    Our findings indicate that the \gpt app ecosystem is still in a nascent stage in providing rich functionalities comparable with its mobile counterparts~\cite{alecci2024revisiting}. 
    It also lacks a mature regulatory mechanism to enforce user privacy compliance and security standards. 
    Our study not only contributes to the improvement of the current store, but also provides insights into the future development of the entire ecosystem.  
    
\end{itemize}

\section{Characterizing existing \gpt plugins (RQ1)} \label{sec:RQ1}
To answer RQ1, we collect existing plugins and their associated artifacts from the \gpt plugin store, and conduct a characterization to provide a comprehensive overview of their functionality distribution. 

\subsection{Plugin Collection}

\paragraph{Plugin Metadata Collection}\label{sec:crawl}
We utilize a web scraper called \emph{Easy Web Data Scraper}~\cite{easy2023web} from the Chrome Web Store to automate the data extraction. To initialize the data collection process, we log our testing account into \gpt and navigate to the plugin store.  
After specifying the representation of an arbitrary GPT plugin UI window, the scraper can automatically identify all the plugin windows on one page. Next, we pinpoint the location of the ``Next Page'' button and establish the crawling speed for individual pages, allowing the scraper to navigate through all accessible pages in the \gpt plugin store and gather window data for each publicly available plugin. 
Consequently, we collect the metadata for each plugin, encompassing the plugin's logo, name, description, legal documents, and email.
Our collection of extention metadata from the GPT store is detailed in Section~\ref{sec:gptstore}. 

\paragraph{Dataset}
Using this crawling method, we construct a longitudinal monitoring dataset for the \gpt plugin store. This monitoring process started after the plugin store ceased accepting new plugin registrations in November 2023, and continued until the \gpt plugin store was no longer open to users, spanning four months until April 2024. 

\subsection{Plugin Category Definitions} \label{sec:category1}
To comprehensively cover all categories of plugins, we first refer to the categorization of three application stores with the highest market share, the Google Play Store, Apple App Store, and Amazon Appstore, respectively.
Specifically, the Google Play Store offers 33 categories~\cite{google2023categories}, the Apple App Store features 27 categories~\cite{apple2023categories}, and the Amazon Appstore encompasses 28 categories~\cite{amazon2023categories}. 
As a result, we identify 40 different categories from them after excluding duplicates.
Note that due to the uniqueness of \gpt plugin functionalities, these app stores' categorization can not be directly applied.  
Therefore, we organize a review group to explore a categorization specifically tailored for the \gpt plugin store.

Our review group consists of three co-authors and four colleagues in our research institute, each with a strong foundation in software engineering and practical experience with \gpt plugins. Within this group, two members possess expertise in software development and plugin creation. We equip them with a specially designed tutorial that succinctly captures the features of \gpt plugins. This preparation paves the way for an informed decision-making process, where the final categorization schema for the \gpt plugin store emerges from a majority vote among these informed group members. Our study has been guided by an ethics committee member of our institute.

The selection process begins with an initial set of 40 app categories. In the first step, group members independently mark those categories that do not apply to the context of LLM apps for further discussion. This process excludes the ``Communication'' category, which predominantly covers instant messaging and voice-over-IP services. The next step involves excluding categories that are either too broad or can be combined with others to prevent classification ambiguity. Consequently, the categories ``Personalization'' and ``Customization'', which lack precise definitions, are removed. Then, the team reviews around 200 randomly selected plugins (20\% of the total) to identify categories specific to \gpt. This results in the discovery of the ``Law'' and ``Plugin Tips'' categories. 
As a result, we ultimately define 21 categories for the \gpt plugin store.
Our categorization achieves an excellent Fleiss' Kappa score of 0.92.

\subsection{Classification Methodology} \label{sec:category2}
After determining the scope of the categories,
we employ the \emph{BART-large-mnli}~\cite{facebook2023bart-large-mnli}, a model developed by Facebook (AI at Meta) for classification of \gpt plugins. This model represents a checkpoint of the BART-large model trained on the MultiNLI (MNLI) dataset, which comprises 433,000 sentence pairs annotated with textual entailment information~\cite{N18-1101}. 
The BART-large-mnli model can be utilized for zero-shot text classification. This is achieved by presenting the text to be classified as the NLI premise and constructing a hypothesis for each potential label~\cite{yin2019zero-shot}. We set three hypotheses for each category, including a positive hypothesis, a contradiction hypothesis, and an irrelevant hypothesis. For example, to determine whether a text sequence pertains to the label ``Games'', a positive hypothesis such as ``This text describes a game'' can be constructed. 
 Subsequently, the probabilities of entailment regarding the hypothesis are translated into probabilities for the corresponding potential label. 

We assign 21 pre-defined categories (Section~\ref{sec:category1}) as the classification labels for our model, and then feed the descriptions of the plugins into the model, expecting it to determine the appropriate category for each plugin. The decision to utilize single-label classification is because plugins tend to offer specialized, singular functionalities, in contrast to mobile apps. We interact with the model via the Inference API provided by Hugging Face~\cite{hugging2023AI}.

\paragraph{Benchmarking} To examine the classification model's performance, we construct a benchmark for our study. We randomly select 100 plugins and apply the classification model to classify them. 
We manually establish the ground truth for these plugins. To avoid bias, three members in our review group are involved in annotating the benchmark, allowing only one category to be assigned per plugin. To determine the category of a plugin, we employ a majority vote mechanism. If the three members assign three different categories to the same plugin, a fourth member is involved to participate in the decision-making process. 
The evaluation of the model's performance adopts a macro-F1 score, which provides a balanced assessment by considering the F1 score across all categories.
This approach highlights the classification model's capability to accurately identify all categories, including those with fewer examples.
After labeling, we achieve an accuracy of 83.3\% and a macro-F1 score of 0.91 for our model.

\begin{figure}[t]
    \centering
    \vspace{-0.1cm}
    \includegraphics[width=0.5\textwidth]{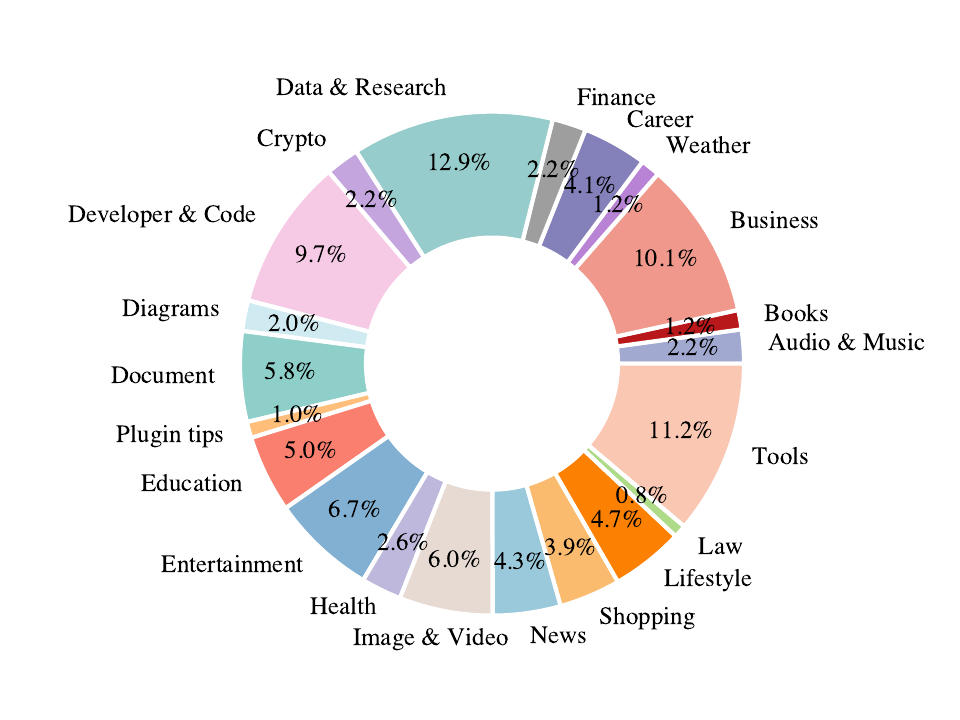}
    \vspace{-0.5cm}
    \caption{Distribution of the number of plugins for the 21 categories}
    \label{fig:plugin_distribution}
     \vspace{-0.5cm}
\end{figure}

\begin{figure*}[t]
    \centering
    \includegraphics[width=1\textwidth,trim={4.5cm 2cm 3.5cm 1.7cm},clip]{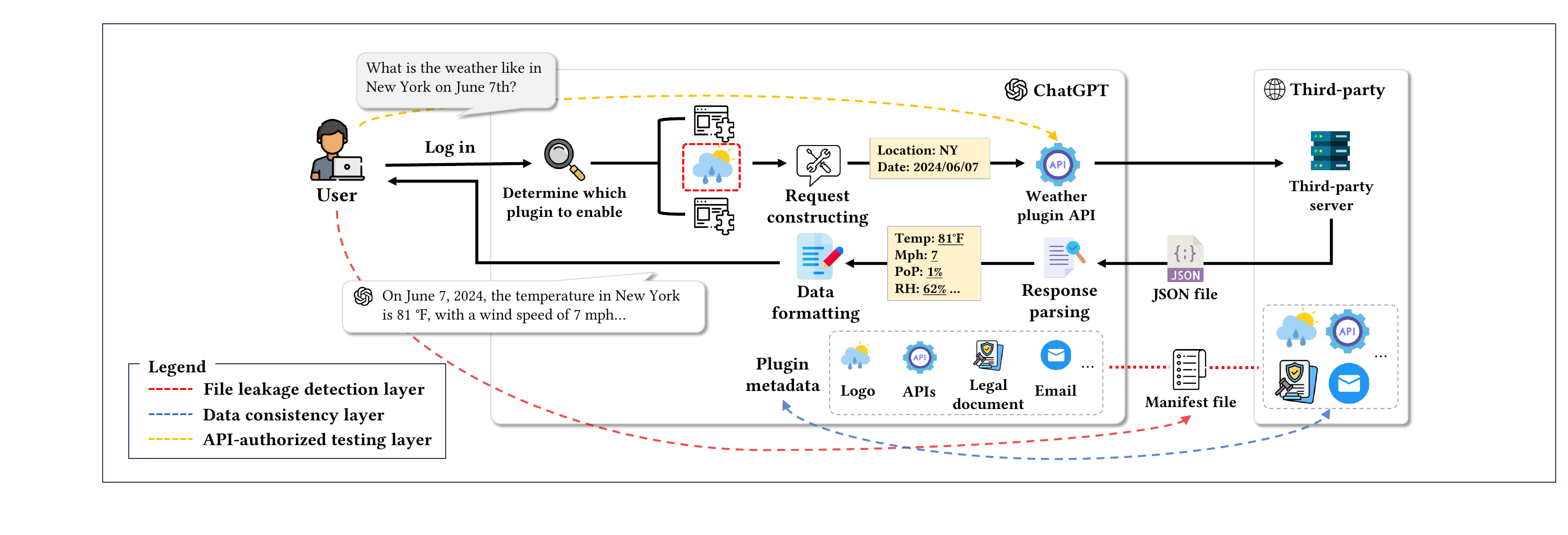}
    \vspace{-0.7cm}
    \caption{The workflow of security assessment model based on the plugin operating mechanism}
    \vspace{-0.3cm}
    \label{fig:operating_mec}
\end{figure*}

\subsection{Distribution of Plugins}

After benchmarking, we apply the classification model to all collected plugins. 
Figure~\ref{fig:plugin_distribution} shows the outcome, which reveals the current status of plugin development and usage categories. More specifically, the category with the highest percentage is ``Data \& Research'', accounting for 12.9\% of the plugins. This underscores the significant interest in utilizing \gpt for data analysis, research purposes, and perhaps data-driven decision-making processes. Following closely is ``Tools'' at 11.2\% and ``Business'' at 10.1\%, suggesting that \gpt plugins are heavily leaned towards productivity and professional use cases and provide functionalities that streamline workflows, enhance business operation, and offer various utilities.

The category ``Developer \& Code'' reserves 9.7\% of the plugins, highlighting the advantage of \gpt in understanding programming and software development contexts. This could encompass a wide range of utilities, from code generation and debugging to more sophisticated uses like automated programming assistance and code optimization. The ``Entertainment'' and ``Image \& Video'' categories, accounting for 6.7\% and 6.0\% respectively, illustrate the use of AI in creating engaging content, editing images, and producing videos, reflecting the growing intersection of technology and creativity.

Conversely, the \gpt specific plugin category ``Law'' accounts for 0.8\% of the total. Although this proportion might seem small, indicating that the development in this area is still in its initial stages, this phenomenon reflects how AI is gradually taking over complex and professional tasks typically requiring direct human involvement, such as legal interpretation or case analysis.

\paragraph{Country/Region} We discover that certain plugins are tailored to serve specific countries and regions. For instance, the primary function of the plugin ``Search UK Companies'' is to fetch public information on UK-registered companies and their officers from the companies' houses. Hence, we employ the \texttt{en\_core\_web\_sm} model within the Spacy~\cite{spacy} framework to enable automatic identification and matching of countries and regions. The outcomes are presented in Table~\ref{tab:plugin_country}. 
Surprisingly, the number of plugins for Japan exceeds those for the US, where \oai is registered, followed by the UK, Australia, South Korea, and Canada. These plugins typically provide information related to the user's geographical location, such as real estate sales and job searches, indirectly assuming the collection of the user's geographical data.
We remark that \oai does not restrict the availability of plugins in different countries and regions. On the other hand, \oai has not enforced region-specific regulatory compliance, such as GDPR and CCPA, during the plugin development.
\begin{formal}
\paragraph{Finding 1}
We categorize all plugins available in the plugin store into 21 different categories, with most falling into practical categories like ``Data \& Research'' and ``Tools''. Additionally, we observe that these plugins are not restricted by country or region, which can potentially lead to issues related to data regulation compliance and other privacy concerns.
\end{formal}

\begin{table}[t]
\caption{The distribution of country-specific plugins}
\footnotesize
\label{tab:plugin_country}
\resizebox{1\linewidth}{!}{
\begin{tabular}{c|c|c|c}
\hline

\hline \textbf{Country \& Region} & \textbf{Plugin Number} & \textbf{Country \& Region} & \textbf{Plugin Number}\\ \hline
     Japan & 27& USA & 24\\ \hline
     UK & 8 & Australia & 6\\ \hline
     Korea & 5 & Canada & 4\\ \hline
     Germany & 4 & Singapore & 4\\ \hline
     China & 3 & Switzerland & 3\\ \hline
     Austria &  1 & Brazil &  1\\ \hline
     India &  1 & Ireland &  1\\ \hline
     Israel &  1 & Italy &  1\\ \hline
     Taiwan&  1 & Portugal &  1\\ \hline
     Turkey &  1 & Netherlands & 1\\ \hline
     
\hline
\end{tabular}}
\vspace{-0.5cm}
\end{table}

\section{Understanding App Deployment and Runtime Execution (RQ2)} \label{sec:RQ2}

\gpt plugins employ an innovative interaction mode specifically designed to augment and elevate the functionalities of \gpt. 
It serves as a bridge that allows developers to contribute their creativity and expertise. By creating and publishing diverse plugins, developers can integrate additional features into \gpt, thereby significantly enriching the user experience. 
These plugins can range from enhancing language processing capabilities to introducing novel interactive tools, all crafted to provide users with a more versatile, customized, and engaging interaction with \gpt. 

We develop a test app to study the deployment and execution mechanisms of \gpt plugins. 
Following the official development documentation~\cite{chatgptmanifest}, we initialize and complete the test app, which includes a simple API supported using Django~\cite{django}. 
We implement a Google authentication to log into our test app. 
During the analysis, we invoke and interact with this test app using our test account. 
Simultaneously, we record all requests to the Django server and use Fiddler~\cite{fiddler} to capture the communication packets. 
Below, we reveal the plugin deployment and execution mechanisms learned through our analysis. 

\paragraph{Deployment of a Plugin}
Initially, for the third-party developer who aim to feature their plugin in the \gpt plugin store, it is mandatory to provide \oai with a manifest file. This crucial document should encompass a comprehensive range of metadata about the plugin, such as its name, logo, legal documentation, APIs, plugin description, and email address. 
Importantly, the manifest file is a pre-requisite and needs to be accessible from the API's domain, specifically located at the path ``\texttt{/.well-known/ai-plugin.json}''.
In addition, \oai clearly specifies which data in this document must be presented to users, such as plugin name, description, and legal documentation. 
\oai also stipulates which data must not be disclosed to users, such as API information and authentication settings (to be detailed in Section~\ref{sec:def}).

\paragraph{Plugin Workflow}
Figure~\ref{fig:operating_mec} illustrates the operational logic of plugins. 
Upon successfully deploying a plugin, users are free to install it in their \gpt environment. 
To use it, the user needs to input a prompt that drives \gpt to seek support from the installed plugin. 
\gpt first evaluates the user's prompt and uses the description from the plugin's submitted manifest file to determine which plugin should be activated. Once a selection is made, \gpt constructs a request by extracting data from the prompt based on the plugin's API. This request is then sent to the third-party server via the API, which responds with a JSON data packet. \gpt analyzes the data from this packet, formats it, and returns it to the user in natural language. This completes the entire process of querying information in \gpt using a plugin. 

For example, the user can start it by simply typing a prompt \emph{``What is the weather like in New York on June 7th?''}. 
Upon receiving the prompt, \gpt identifies and activates the plugin named ``Weather Manager'' as the suitable choice for this query. It then extracts two pieces of information from the prompt, the location and the date, which are passed as parameters to the API in the request. The server of ``Weather Manager'' responds with a JSON data packet, which includes details about the temperature, wind speed, humidity, and the probability of precipitation in New York on June 7. Finally, \gpt processes the data and returns it to the user in natural language like \emph{``On June 7, 2024, the temperature in New York is 81°F, with a wind speed of 7 mph...''}.

\begin{formal}
\paragraph{Finding 2}
Third-party plugins deployed on \gpt are integrated with the platform through a manifest file. This file acts as the \emph{bridge} between the plugin and the \gpt platform. It defines various metadata and functionalities of the plugin, and enables \gpt to recognize, load, and correctly execute the plugin's features. Given that these plugins handle large amounts of user data transmitted through the LLM, they must adhere to strict data protection protocols and security standards.
\end{formal}
\section{Assessing Security and Privacy (RQ3)}\label{sec:exposure}
The core idea of our study is to evaluate the plugin's data exposure behavior, which may lead to potential security risks (as discussed in Section~\ref{sec:broad}). In response to the special interaction mode, we first define five representations based on the elements involved in user interactions with the plugin. Then, we propose a three-layer assessment model to define five types of data exposures and develop three progressive methods to assess them.

\subsection{Exposure Definition}\label{sec:def}

\paragraph{Problem Definition}
In our definition, we initially categorize the fields of the manifest file. $O_{r}$ represents the set of fields that OpenAI requires third-party developers to include in the plugin specification documentation~\cite{chatgptmanifest}. $O_{p}$ signifies the subset of these fields that must be disclosed to users, where $O_{p} \subseteq O_{r}$. The set $O_{h} = O_{r} - O_{p}$ denotes the collection of fields that cannot be disclosed to users. 

\paragraph{Definition 1: Plugin metadata representation ($\mathcal{U}$)} The public data exposed to users on the \gpt plugin store interface is represented as a set of tuples $\mathcal{U} = \{u |\,u : (n_{u}, l_{u}, d_{u}) \}$, each $u$ represents the interface data of a plugin, where $n_{u}$ stands for the plugin's interface name, $l_{u}$ represents the link of the plugin's legal document, and $d_{u}$ denotes the plugin's description on the interface.

\paragraph{Definition 2: manifest data representation ($\mathcal{M}$)} $\mathcal{M}$ represents the set of all manifest files for each plugin. $\mathcal{M} = \{m |\,m : (n_{m}, l_{m}, k_{m},\\ h_{m}, t_{m}, d_{m}), \neg h_{m} \Rightarrow t_{m} = \emptyset \}$. $h_{m}$ specifies whether a plugin requires multi-authentication, with $\neg h_{m}$ indicating not required. $n_{m}$ denotes the plugin name for users. $l_{m}$ represents the link to the plugin's legal document in the manifest file, $k_{m}$ is the plugin API link, and $t_{m}$ corresponds to \oai API token. $d_{m}$ describes the plugin in the manifest file. Notably, a token exists only when the $h_{m}$ is true.

\paragraph{Definition 3: APIs representation ($\mathcal{A}$)} $\mathcal{A}$ denotes the collection of all APIs within plugins. $\mathcal{A} = \{a |\,a : \{(i_{1}, r_{1}), (i_{2}, r_{2}),...\} \}$, where $a$ represents all the APIs involved in a plugin and their response data, stored in the form of key-value pairs. $i_{n}$ denotes API, and $r_{n}$ indicates the response data of the corresponding API.

\paragraph{Definition 4: Plugin representation ($\mathcal{P}$)} We designate $\mathcal{P}$ represent all the plugins present in the \gpt plugin store. $\mathcal{P} = \{p |\,p : (u,m,a), u\in \mathcal{U}, m\in \mathcal{M}, a\in \mathcal{A}, u\neq \emptyset \}$.

\paragraph{Definition 5: Invalid response ($\mathcal{N}$)} Even if an API successfully responds to a request (with a status code of 200), it does not necessarily mean that the API has returned meaningful information. For instance, an API might return ``Service Unavailable'', ``Server error'', or ``No requested privileges'' when requested without token authentication. We define the set of such meaningless return results as $\mathcal{N}$.

Below, we introduce each exposure within each layer.
\paragraph{File leakage detection layer}
In the first layer, we primarily analyze the security of plugins from the user's perspective. According to the regulations of the \oai plugin store ecosystem, the only information about plugins that users can obtain is the plugin's metadata. The sole method of interaction with the plugin is through entering prompts in ChatGPT. Based on this principle, we establish \emph{Exposure 1}.

\textbf{\emph{Exposure 1: Non-Empty Manifests}} The non-emptiness of a plugin's manifest file $m$ signifies that users can retrieve the configuration data that third parties have supplied to OpenAI.
\begin{equation}
\label{equ:1}
    \exists p \in \mathcal{P}, m\neq \emptyset.
\end{equation}

\paragraph{Data consistency layer} 
At this layer, we assess the accuracy of the configuration information provided to \oai by plugin developers from a system perspective. Hence, we define \emph{Exposure 2}.

\textbf{\emph{Exposure 2: Discrepancies Metadata}} The metadata provided by third parties to users differs from what is submitted to OpenAI, including the name, description, and legal document link of the plugin.
\begin{multline}
\label{equ:2}
\exists p \in \mathcal{P},
\left(\mathrm{sim}(n_{u},n_{m}) < \vartheta_{1}\right)  
\wedge \\ \left(\mathrm{sim}(d_{u},d_{m}) < \vartheta_{2}\right) \vee \left(\mathrm{sim}(l_{u},l_{m}) < \vartheta_{3}\right)
\end{multline}

\paragraph{API-authorized testing layer} 
We conduct unauthorized external access tests on the plugin's API. From a third-party perspective, if external entities can directly invoke the API provided to \oai, it could lead to unauthorized data access and data leaks on the third-party server. Therefore, we define \emph{Exposure 3} as an authentication method based solely on \oai account login ($h_{m}$ = true). Authentication methods that involve additional account logins are defined as \emph{Exposure 4}, and methods that utilize an \oai token for internal authentication are defined as \emph{Exposure 5}.

\textbf{\emph{Exposure 3: Single Authentication External API Calls.}} Through the API links ($k_{m}$) in the manifest file, all API paths and their parameter settings within the plugin can be discovered. This allows for the construction of requests outside \oai and the sending of these requests to the APIs, which can respond normally and return valid data.
\begin{equation}
\begin{aligned}
\label{equ:3}
    \exists p \in \mathcal{P}, &\quad k_{m} \neq \emptyset \vee \neg h_{m}, \\ \exists a \in \mathcal{A}, \forall x, &\quad r_{x} \neq \emptyset \wedge r_{x} \notin \mathcal{N}.
\end{aligned}
\end{equation}

\textbf{\emph{Exposure 4: Multi-Authentication External API Calls.}} When the auth ($h_{m}$) is true, it means multi-authentication is required to access the API. However, under this exposure condition, the API can still return valid data from outside OpenAI.
\begin{equation}
\begin{aligned}
\label{equ:4}
    \exists p \in \mathcal{P}, k_{m} \neq \emptyset \vee h_{m}, \\
    \exists a \in \mathcal{A}, \forall x, r_{x} \neq \emptyset \wedge r_{x} \notin \mathcal{N}.
\end{aligned}
\end{equation}

\textbf{\emph{Exposure 5: Token Leakage.}} Even if an API cannot be accessed from outside OpenAI, users can still obtain access to the API by including a leaked token ($t_{m}$) from the manifest file in their request, thereby obtaining valid information.
\begin{equation}
\begin{aligned}
\label{equ:5}
    \exists p \in \mathcal{P}, k_{m}, t_{m} \neq \emptyset \vee h_{m}, \\
    \exists a \in \mathcal{A}, \forall x, r_{x} \neq \emptyset \wedge r_{x} \notin \mathcal{N}.
\end{aligned}
\end{equation}

\paragraph{Discussion} 
All the exposures we have defined are based on the \textbf{file leakage detection layer}. This layer most directly reveals the plugin's exposure behavior, as it is used to detect situations where users obtain data they should not have access to (\emph{Exposure 1}). Specifically, \textbf{Data consistency layer} emerges from checking the consistency between user-facing and system-facing data (\emph{Exposure 2}). \textbf{API-authorized testing layer} pertains to detecting broken access control vulnerabilities related to the API (\emph{Exposure 3}, \emph{Exposure 4}, and \emph{Exposure 5}).

\begin{table}[t]
\caption{Plugin manifest file accessibility distribution\label{tab:manifest_access}}
\vspace{-0.2cm}
\small
\begin{tabular}{c|c c c c}
\hline

\hline   \multicolumn{1}{c|}{ \textbf{Accessible} } & \multicolumn{4}{c}{ \textbf{Inaccessible} }\\ \cline{1-5}

  \multicolumn{1}{c|}{Unrelated} & \multicolumn{1}{c|}{Timeout} & \multicolumn{1}{c|}{Google drive} & \multicolumn{1}{c|}{Github} & \multicolumn{1}{c}{Server error} \\ \hline
 
 104 &\multicolumn{1}{c|}{16} &\multicolumn{1}{c|}{6} &\multicolumn{1}{c|}{19} &\multicolumn{1}{c}{518} \\ \hline
 
 \hline

\end{tabular}
\vspace{-0.2cm}
\end{table}

\begin{figure}[t]
    \centering
    \includegraphics[width=0.47\textwidth]{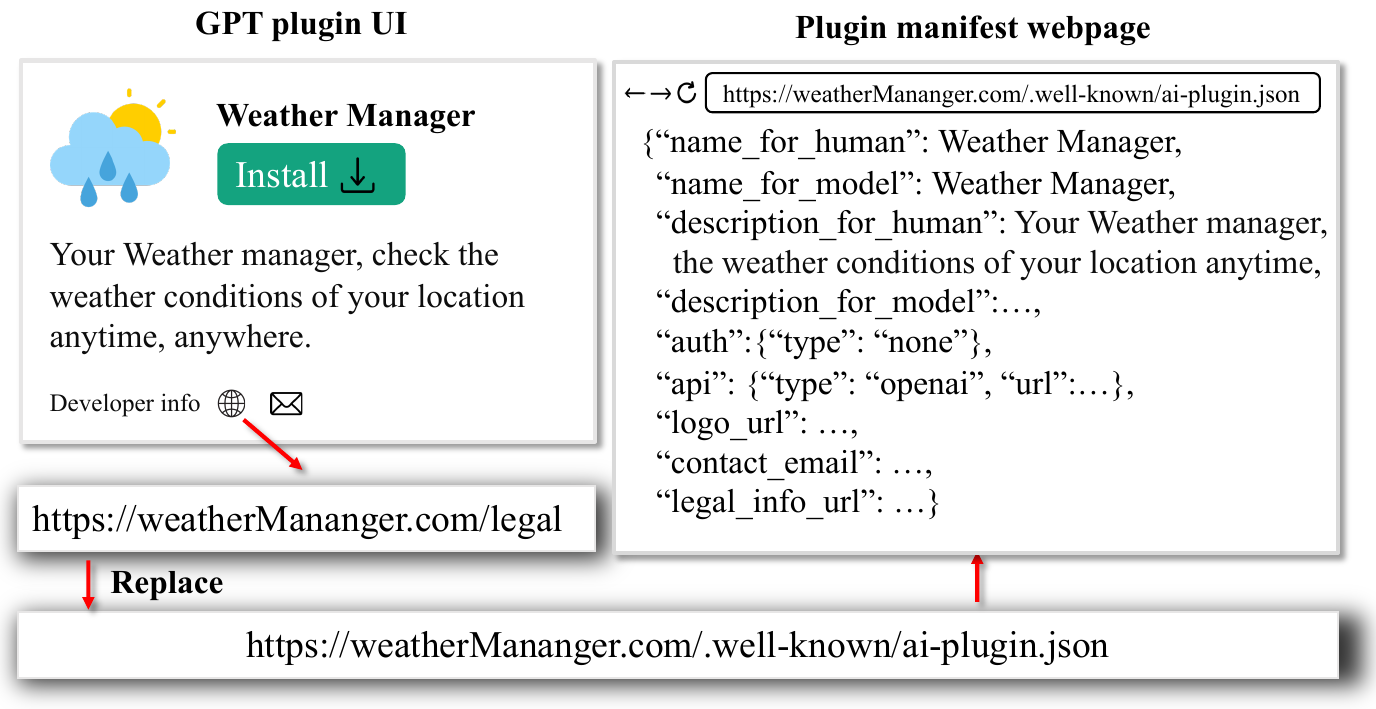}
    \vspace{-0.2cm}
    \caption{The process of getting the plugin manifest file}
    \label{fig:get_manifest}
    \vspace{-0.4cm}
\end{figure}

\subsection{Assessment of File Leakage}\label{sec:pmf}

As mentioned in Section~\ref{sec:RQ2}, each plugin is required to provide a manifest file hosted at a specified path under the API domain. Figure~\ref{fig:get_manifest} introduces how we obtain the manifest file of a plugin in \textbf{File leakage detection layer}.
More specifically, we first locate the interactive page of a plugin in the \gpt plugin store, then click on the ``globe'' icon within the developer info to acquire the URL of the legal document. Finally, we replace the outermost path of the URL with ``\texttt{/.well-known/ai-plugin.json}'' or ``\texttt{/.well-known}'' and attempt to access it. If the URL can be accessed successfully, the returned results are displayed on the plugin manifest webpage shown in Figure~\ref{fig:get_manifest} (right). We utilize BeautifulSoup~\cite{BeautifulSoup} to create a script that automates the process of modifying the plugin's legal document URL and crawling the plugin manifest file.

\paragraph{Evaluation} To evaluate our crawling method, we separately record the status code for accessible and inaccessible URLs. Among the URLs that could not be accessed normally, 16 returned \texttt{timeout}, possibly due to the website identifying script behavior and choosing not to respond. Therefore, we manually visit these URLs, and find that within them 3 are able to respond normally. We also discover that the legal documents of 25 plugins use links for open Google driver and Github, which hinders our ability to access the manifest files through alterations in the path.

On the other hand, we discover that even if the URL status code is 200, it is possible that the content returned is unrelated to the plugin's configuration. Therefore, we filter the URLs by setting seed words such as ``auth'', ``api'', ``legal\_info\_url'' etc., ultimately identifying 104 unrelated to the manifest file. The details are shown in Table~\ref{tab:manifest_access}. 

\begin{formal}
\paragraph{Finding 3}
We totally detect 368 (35.7\%) plugins that leak manifest files, which enables external access to the plugin’s configuration settings, including sensitive data like third-party APIs designed exclusively for \oai.
\end{formal}

\subsection{Assessment of Data Similarity}
The legal document provided to \oai might differ from the URL given to users. This discrepancy can lead to confusion as users may not have access to the same terms and conditions or privacy policies that \oai has reviewed. Similarly, to improve the ranking of the plugin (since the \gpt plugin store defaults to alphabetical ordering), developers might present the plugin name to users as ``weather manager'', while declaring it in the manifest file as ``AAA\_weather\_manager''. Furthermore, \gpt matches the user's prompt to the plugin's description to trigger the corresponding plugin. If the  $d_{u}$ and $d_{m}$ do not align, it may create opportunities for malicious plugins. Therefore, in the \textbf{Data consistency layer}, we precisely check such deceptive practices aimed at misleading both the system and the users.

In the absence of training data, we adopt the cosine similarity~\cite{rahutomo2012semantic} method to accurately and efficiently compare the consistency of data provided by plugins. A plugin is considered to provide inconsistent data when the cosine similarity between these pieces of data falls below a preset threshold. Considering the special characters in the plugin name and the model's semantic understanding, we have set $\vartheta_{1}$ and $\vartheta_{2}$ to 0.85 and 0.8 (empirically set), respectively. $\vartheta_{3}$ is set to 1, as it is assumed that $l_{u}$ and $l_{m}$ must be the same.

\paragraph{Evaluation}
Given the absence of a benchmark in the existing literature, we take the initiative to create one for our data similarity study. we randomly selected 30 plugins from 368 with leaked manifest files. Among these, we identify inconsistencies in the names of 4 plugins and in the description of 2 plugins, both $\vartheta_{1}$ and $\vartheta_{2}$ accurately exclude these cases. Furthermore, given the possibility that different URLs may lead to the same web page content, we also perform a consistency check on these web pages and do not find such instances.

\begin{formal}
\paragraph{Finding 4}
We have identified a total of 69 plugins where the metadata submitted to \oai does not match the metadata provided to users. Among these, the names of 34 plugins are inconsistent, 8 descriptions differ, and 27 legal document URLs do not match. This discrepancy either arises from developers failing to update their plugins promptly or from developers intentionally providing false metadata to deceive both \oai and users.
\end{formal}

\begin{figure}[t]
    \centering
    \includegraphics[width=0.475\textwidth]{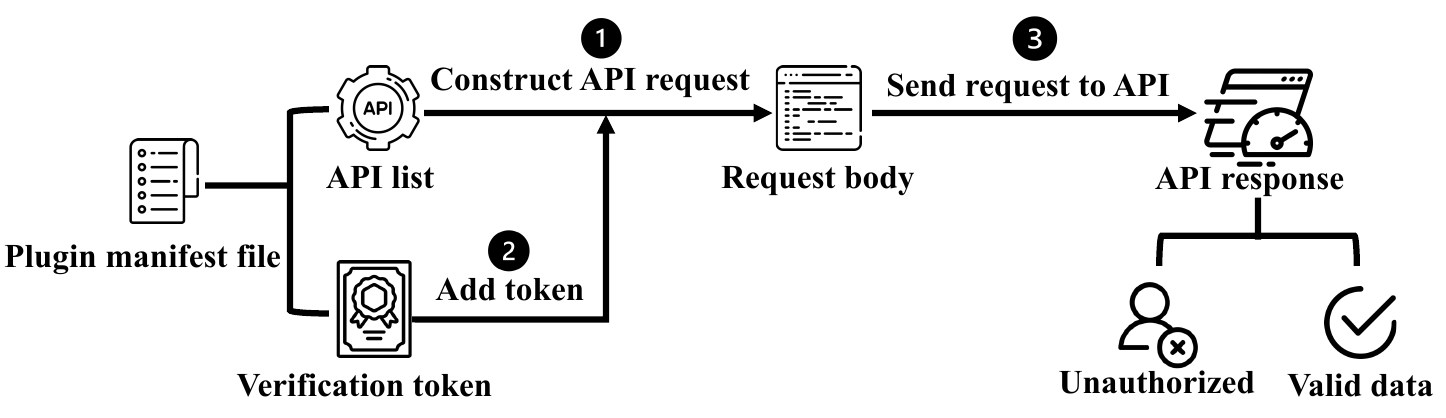}
    \vspace{-0.4cm}
    \caption{Plugin APIs request flow}
    \label{fig:api_request_flow}
\end{figure}

\subsection{Construction of the Plugin's API Request}

Figure~\ref{fig:api_request_flow} illustrates the process of verifying the external accessibility of APIs at the \textbf{API-authorized testing layer}. Initially, we utilize the API link ($k_{m}$) leaked in the manifest file to identify all relevant APIs (API list). Then, we construct an API request body for each API based on their declared parameters (step \ding{182} in Figure~\ref{fig:api_request_flow}). For APIs that declare a verification token, we include their token in the request body to validate its effectiveness (step \ding{183}). Subsequently, we use the Request library~\cite{request231} of Python to send the request body to the third-party API server and receive a response (step \ding{184}). If the API returns valid information successfully, it indicates that the API has failed the external authentication request test. Conversely, if the API rejects the external request, it passes the test.

\paragraph{Evaluation} 
Considering that third-party servers might impose restrictions on API requests and potential network issues with local requests, we employ three different IPs to simultaneously send requests to the APIs during the different periods. 

For APIs from which we could not obtain valid information, we attempt multiple requests to ensure robustness and minimize the impact of any anomalies. Aside from responses returning ``unauthorized'', which remain unchanged, we find that API responses can vary with changes in time and IP, introducing uncertainty into the results. For example, accessing the API at $\alpha$ time with IP $A$ returns a status code of 404, while accessing it at $\beta$ returns a status code of 200. Detailed data is shown in Table~\ref{tab:api_access}.

\begin{table}[t]
\setlength{\tabcolsep}{2pt}
\caption{The distribution of API response\label{tab:api_access}}
\vspace{-0.2cm}
\resizebox{1\linewidth}{!}{%
\begin{tabular}{c|c c c|c c c c}
\hline

\hline \multirow{2}{*}{\shortstack{\textbf{API} \\ \textbf{responsiveness}}} & \multicolumn{3}{c|}{\textbf{Auth types}} & \multicolumn{4}{c}{\textbf{Reasons}} \\ \cline{2-8}

& \multicolumn{1}{c|}{\texttt{none}} & \multicolumn{1}{c|}{\texttt{others}} & \multicolumn{1}{c|}{Token$\dagger$} & \multicolumn{1}{c|}{Change$\ddagger$} & \multicolumn{1}{c|}{Unauthorized} & \multicolumn{1}{c|}{\makecell[ct]{Client\\errors}}  
   & \multicolumn{1}{c}{\makecell[ct]{Rate\\limiting}}\\ \hline
 
respondable &\multicolumn{1}{c|}{141} &\multicolumn{1}{c|}{32} &\multicolumn{1}{c|}{8} &\multicolumn{1}{c|}{5} &\multicolumn{1}{c|}{-}
 &\multicolumn{1}{c|}{-}
 &\multicolumn{1}{c}{-}\\ \hline
 
non-respondable &\multicolumn{1}{c|}{87} &\multicolumn{1}{c|}{72} &\multicolumn{1}{c|}{-} &\multicolumn{1}{c|}{-}
 &\multicolumn{1}{c|}{55}
 &\multicolumn{1}{c|}{62}
 &\multicolumn{1}{c}{42} \\ \hline
 
 \hline
\end{tabular}}

\begin{flushleft}
    \footnotesize
    \textsuperscript{$\dagger$} Token: Include verification token in the API request body.\\
    \textsuperscript{$\ddagger$} Change: Manually requestable.
\end{flushleft}
\vspace{-0.5cm}
\end{table}

\begin{formal}
\paragraph{Finding 5}
173~(52.1\%) plugins can retrieve valid information from their specialized API for \oai, including 141 with an $h_{m}$ is true.
For plugins with authentication requirements, 24 plugins can return valid information, while 8 plugins are able to retrieve valid information after adding verification \oai token.
In addition, 159 plugins fail to return valid information successfully. Among them, 55 plugins lack authorization information; 62 plugins face client errors, such as access prohibited and resource not found; 42 plugins cannot be successfully requested because of rate limiting.
\end{formal}

\subsection{Plugin Legal Documents}
As the only document available to users, the plugin's legal document serves as the sole channel for users to understand the risks and security of the plugin. However, \oai has not stipulated or reviewed the contents of this document, resulting in third-party developers potentially linking to web pages irrelevant to the plugin or suboptimal content. Hence, we evaluate the content of the legal documents provided by the plugin.

The process begins with an initial screening of the documents. Based on the work of several previous studies of legal documents~\cite{wilson2016creation,soneji2024demystifying,chen2023investigating}, we develop a seed word library for legal attributes, which is presented in Table~\ref{tab:seed}. This library is used to search through all documents, and we consider a document to have legal attributes if its content contains any seed word from the library (except the navigation bar). Although this method may result in some false positives, it ensures that all documents containing legal attributes are identified. Previous studies~\cite{li2019filtering,xie2022scrutinizing,yan2024quality} have demonstrated that the seed word library can efficiently filter out the target documents with an accuracy up to 98\%. 
The results indicate that among all documents, 767 (73.9\%) can be accessed programmatically, and only 649 (62.5\%) of the total documents are identified as containing legal attributes, most of them pertaining to the ``Terms of Service'', ``Privacy Policy'' and ``Legal Information'', with the specific data distribution displayed in Table~\ref{tab:legal_doc}. To ensure the accuracy of the results, two members with legal expertise from the review group conduct a manual inspection. In those documents that do not contain legal attributes, we find that the majority are company websites, while others include Github repositories and portfolio websites. 

\begin{table}[t]
\caption{Legal attributes seed word library\label{tab:seed}}
\small
\vspace{-0.2cm}
\begin{tabular}{p{8cm}}
\hline

\hline Privacy, Regulation, Statute, Provision, Affiliates, Collection, Opt-Out, Personal Information, User Consent, Retention Period, Data Protection, Data Subject, Data Controller, Data Processor, Legitimate Interest, Cross-Border Data Transfers   \\ \hline

\hline
\end{tabular}
\end{table}

\begin{table}[t]
\caption{The accessibility of plugins' legal documents\label{tab:legal_doc}}
\vspace{-0.2cm}
\resizebox{1\linewidth}{!}{%
\begin{tabular}{c|c c c c}
\hline

\hline   \multicolumn{1}{c|}{ \textbf{Inaccessible} } & \multicolumn{4}{c}{ \textbf{Accessible} }\\ \cline{2-5}
\multicolumn{1}{c|}{} & \multicolumn{1}{c|}{Unrelated} & \multicolumn{1}{c|}{Terms of service} & \multicolumn{1}{c|}{Privacy policy} & \multicolumn{1}{c}{Other legal doc} \\ \hline
271 &\multicolumn{1}{c|}{118} &\multicolumn{1}{c|}{391} &\multicolumn{1}{c|}{116} & \multicolumn{1}{c}{142} \\ \hline

\hline
\end{tabular}}
\end{table}

\subsection{Distribution by Plugin Category and Email Domain}

\begin{table*}[t]
\def\arraystretch{0.98}
\setlength{\tabcolsep}{4pt}
\caption{Five types exposures in different plugin categories}
\label{fig:result_category}
\vspace{-0.3cm}
\small
    \begin{tabular}{p{0.24\linewidth}|p{0.12\linewidth}| p{0.13\linewidth}| p{0.12\linewidth}| p{0.12\linewidth}| p{0.12\linewidth}}
    \hline

    \hline
    \multirow{2}{*}{\textbf{Plugin Category}} &\multicolumn{5}{c}{\textbf{Five Types Exposures}} \\ \cline{2-6}
    
     &\textbf{File leakage}&\textbf{Inconsistent data}&\textbf{Single auth}&\textbf{Multi auth}&\textbf{Token auth}\\ \hline
    
    Audio \& Music & 10 \hfill (47.6\%) & 1 \hfill (4.8\%) & 3 \hfill (14.3\%) & 3 \hfill (14.3\%) & 0 \hfill (0.0\%)  \\ \hline
    Books & 2 \hfill (16.7\%) & 0 \hfill (0.0\%) & 1 \hfill (8.3\%) & 1 \hfill (8.3\%) & 0 \hfill (0.0\%)  \\ \hline
    Business & 30 \hfill (30.6\%)& 1 \hfill (1.0\%)& 15 \hfill (15.3\%) & 1 \hfill (1.0\%) & 0 \hfill (0.0\%) \\ \hline
    Career & 21 \hfill (52.5\%)& 1 \hfill (2.5\%)& 10 \hfill (25.0\%) & 0 \hfill (0.0\%) & 1 \hfill (2.5\%)  \\ \hline
    Diagram & 10 \hfill (50.0\%) & 3 \hfill (15.0\%)& 4 \hfill (20.0\%) & 0 \hfill (0.0\%) & 0 \hfill (0.0\%)  \\ \hline
    Crypto & 6 \hfill (26.1\%) & 1 \hfill (4.3\%) & 1 \hfill (4.3\%) & 0 \hfill (0.0\%) &0 \hfill (0.0\%)  \\ \hline
    Data \& Research & 38 \hfill (29.9\%)& 4 \hfill (3.1\%)& 15 \hfill (11.8\%) & 2 \hfill (1.6\%) & 2 \hfill (1.6\%)  \\ \hline
    Developer \& Code & 36 \hfill (37.9\%)& 6 \hfill (6.3\%)& 11 \hfill (11.6\%) & 5 \hfill (5.3\%) & 2 \hfill (2.1\%)  \\ \hline
    Document & 22 \hfill (39.3\%) & 4 \hfill (7.1\%) & 5 \hfill (8.9\%) & 5 \hfill (8.9\%) & 1 \hfill (1.8\%)  \\ \hline
    Education & 18 \hfill (36.7\%) & 2 \hfill (4.1\%) & 9 \hfill (18.4\%) & 1 \hfill (2.0\%) & 0 \hfill (0.0\%)  \\ \hline
    Entertainment & 33 \hfill (42.3\%)& 4 \hfill (5.1\%)& 10 \hfill (12.8\%) & 0 \hfill (0.0\%) & 1 \hfill (1.3\%)  \\ \hline
    Finance & 7 \hfill (30.4\%) & 0 \hfill (0.0\%) & 3 \hfill (13.0\%) & 0 \hfill (0.0\%) & 0 \hfill (0.0\%)  \\ \hline
    Health & 2 \hfill (8.0\%) & 0 \hfill (0.0\%) & 0 \hfill (0.0\%) & 0 \hfill (0.0\%) & 0 \hfill (0.0\%)  \\ \hline
    Image \& Video  & 27 \hfill (45.8\%)& 3 \hfill (5.1\%) & 12 \hfill (20.3\%) & 1 \hfill (1.7\%) & 0 \hfill (0.0\%)  \\ \hline
    Law & 3 \hfill (37.5\%) & 0 \hfill (0.0\%) & 3 \hfill (37.5\%) & 0 \hfill (0.0\%) & 0 \hfill (0.0\%)  \\ \hline
    News & 14 \hfill (33.3\%) & 1 \hfill (2.4\%) & 7 \hfill (16.7\%) & 1 \hfill (2.4\%) & 0 \hfill (0.0\%)  \\ \hline
    Plugin Tips & 4 \hfill (40.0\%) & 2 \hfill (20.0\%)& 2 \hfill (2.1\%) & 0 \hfill (0.0\%) & 0 \hfill (0.0\%) \\ \hline
    Shopping & 18 \hfill (42.9\%)& 3 \hfill (7.1\%)& 11 \hfill (26.2\%) & 0 \hfill (0.0\%) & 0 \hfill (0.0\%)  \\ \hline
    Lifestyle & 18 \hfill (20.0\%) & 1 \hfill (1.1\%) & 8 \hfill (8.9\%) & 0 \hfill (0.0\%) & 0 \hfill (0.0\%)  \\ \hline
    Tools & 44 \hfill (40.4\%)& 12 \hfill (11.0\%)& 10 \hfill (9.2\%) & 4 \hfill (3.7\%) & 1 \hfill (0.9\%)  \\ \hline
    Weather & 5 \hfill (41.7\%) & 1 \hfill (8.3\%) & 1 \hfill (8.3\%) & 0 \hfill (0.0\%) & 0 \hfill (0.0\%)  \\ \hline

    \hline

    \hline
    \end{tabular}

\end{table*}
We categorize the experimental results of five types of exposures according to the plugin category and display them in Table~\ref{fig:result_category}.
We find that file leakage issues are prevalent across all plugin categories. Specifically, the ``Data \& Research'' and ``Developer \& Code'' categories each have 36 instances of file leakage, despite the latter ranking fourth in terms of quantity (9.7\% shown in Figure~\ref{fig:plugin_distribution}). Following are the ``Tools'' and ``Entertainment'' categories, which report 34 and 30 issues, respectively. ``Business'' ranked third in quantity (10.1\%), contains 28 instances. Notably, even though the ``Career'' and ``Diagram'' have fewer plugins, 50\% of them have experienced file leakage. 
Data inconsistencies are mostly found in the ``Tools'' category, likely due to an attempt by plugin developers to improve their ranking in the store by altering their names.

Regarding the API exposure, we have found that it is predominantly distributed across ``Business'', ``Data \& Research'', ``Developer \& Code'', and ``Image \& Video''. These categories also boast a higher quantity of plugins. This implies that a large number of users are likely to encounter third-party API security vulnerabilities when they use LLM.

\begin{figure}[t]
    \centering
    \includegraphics[width=0.38\textwidth]{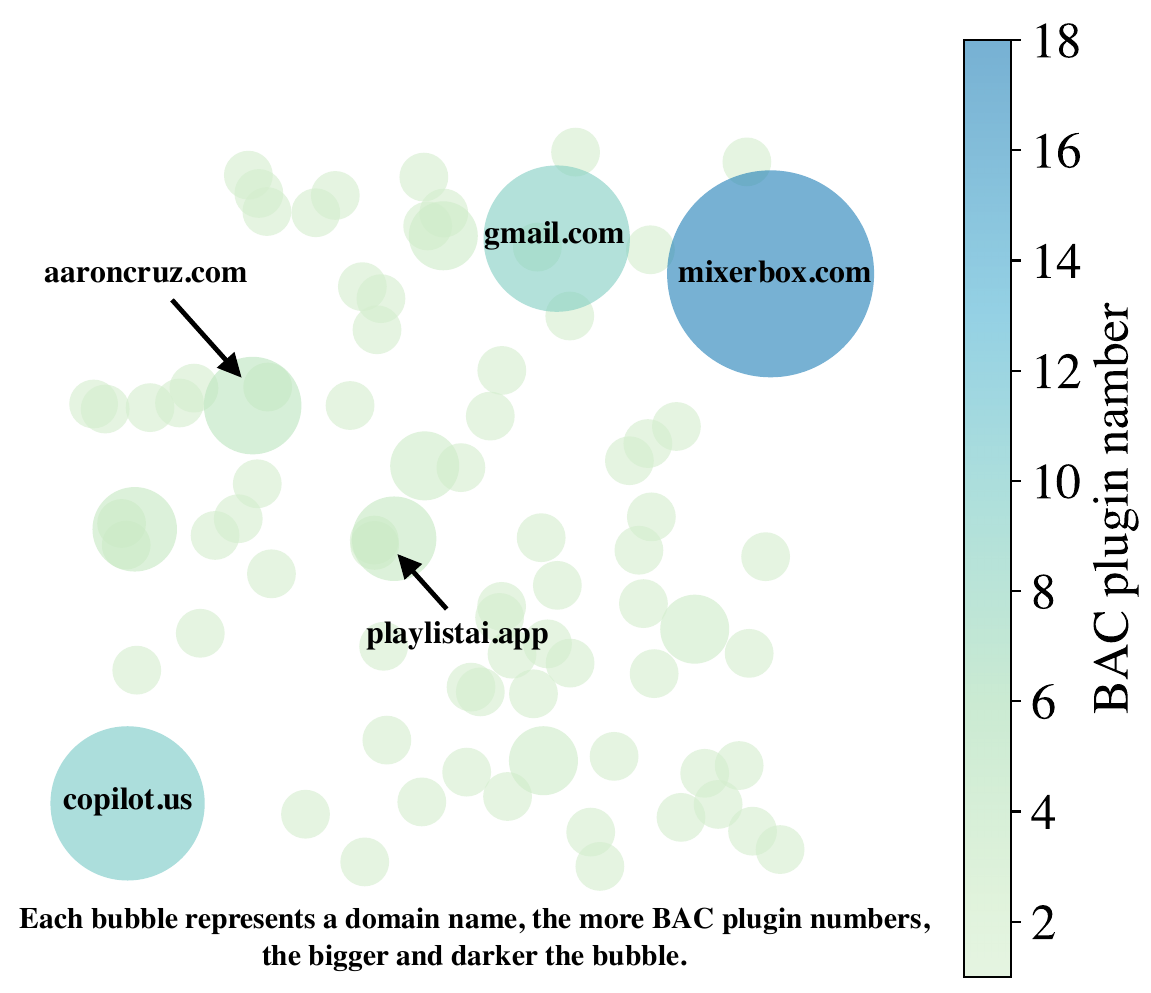}
    \caption{The distribution of developers' email domains for the plugins found to have BAC vulnerabilities}
    \label{fig:plugins_domain}
\end{figure}

We also conduct a statistical analysis on the email domains of plugins with Broken Access Control (BAC) vulnerabilities (\emph{Exposure 3, 4, 5}), as shown in Figure~\ref{fig:plugins_domain}.

\begin{formal}
\paragraph{Finding 6}
Most domains have fewer than 2 BAC plugins. Surprisingly, the ``mixerbox.com'' has 18 BAC plugins, ``copilot.us'' has 10, and ``gmail.com'' has 9. Following closely are ``aaroncruz.com'' and ``playlist.app'' with 4 and 3, respectively. Among these top five domains, all except Gmail are commercial domains. Manual inspection reveals that all these companies are dedicated to developing applications and plugins centered around GPT, offering paid services. This also reflects the current inadequate state of authentication mechanisms employed by these development companies.
\end{formal}

\section{Findings Reporting and Revisit}\label{sec:evo}

\paragraph{Responsible Disclosure}
We have communicated all our findings to \oai through the bugcrowd platform~\cite{bugcrowd2024openai} and actively participated in the resolution process. 
We provided the \oai security team with detailed text-based reproduction steps and test cases and proposed our recommendation to fix the involved issues by incorporating third-party developers. 
The security team of \oai acknowledged our findings but claimed that the involved security issues are out of their scope to resolve.  
Whether they have forwarded our findings to relevant third-party plugin developers remains unknown.

\paragraph{Revisit}
While we were waiting for \oai security team's response, we revisited all plugin manifest files and re-requested all involved APIs on April 9, 2024, which is the last day of \gpt plugin store available. The results are shown in Table~\ref{tab:before_after}.

\begin{table}[t]
\caption{Comparison of five exposures data before and after reporting\label{tab:before_after}}
\small
\resizebox{1\linewidth}{!}{%
\begin{tabular}{c|c c c c c}
\hline

\hline   \multicolumn{1}{c|}{} & \multicolumn{1}{c|}{ \textbf{\makecell[ct]{File\\leakage}} }& \multicolumn{1}{c|}{ \textbf{\makecell[ct]{Inconsistent\\data}} }& \multicolumn{1}{c|}{ \textbf{\makecell[ct]{Single\\auth}}}& \multicolumn{1}{c|}{ \textbf{\makecell[ct]{Multi\\auth}}}& \multicolumn{1}{c}{ \textbf{\makecell[ct]{Token\\auth}}}\\ \cline{1-6}

  \multicolumn{1}{c|}{\textbf{\makecell[ct]{First assessment}}} & \multicolumn{1}{c|}{368} & \multicolumn{1}{c|}{69} & \multicolumn{1}{c|}{141} & \multicolumn{1}{c|}{24} & \multicolumn{1}{c}{8}\\ \hline
 
 \textbf{Revisit (Apr '24)} &\multicolumn{1}{c|}{282} &\multicolumn{1}{c|}{61} &\multicolumn{1}{c|}{89} &\multicolumn{1}{c|}{\textbf{17}}&\multicolumn{1}{c}{5} \\ \hline
 
 \textbf{ Change }&\multicolumn{1}{c|}{\textbf{-23.4\%}} &\multicolumn{1}{c|}{\textbf{-11.6\%}} &\multicolumn{1}{c|}{\textbf{-36.9\%}} &\multicolumn{1}{c|}{\textbf{-29.2\%}}&\multicolumn{1}{c}{\textbf{-37.5\%}} \\ \hline
     
 \hline

\end{tabular}}
\vspace{-0.4cm}
\end{table}
Following the report, we have observed partial resolutions to these security exposures. 
The manifest files of 23.4\% of plugins are now inaccessible to users, and 11.6\% of plugins have addressed issues with data inconsistencies. Among APIs with single and multi-authorization, 36.9\%, and 29.2\%, respectively, can no longer make external requests. It is worth noting that out of 8 APIs that previously permitted data access via tokens, 5 are still accessible.

\section{Discussion}
Our results reveal that the integration of third-party services within large language model platforms, while greatly enhancing functionality and flexibility, and improving user experience and processing efficiency, still introduces significant data security risks. 
These risks may occur in other software with similar architectures (e.g., the broken access control, a frequently occurring vulnerability~\cite{owasp2021top}). Some risks are specific to features unique to \gpt, such as the manifest files required by plugins.
In this section, we primarily introduce 3 potential risks caused by the security exposures discussed in Section~\ref{sec:exposure}~(Section~\ref{sec:broad}). Next, We conduct a study on the legacy plugins in the current GPT Store (Section~\ref{sec:gptstore}). We then offer recommendations to \oai and third-party developers to jointly maintain the security of GPT app ecosystem~(Section~\ref{sec:recommendations}), and also discuss the ecosystem of other LLM-based third-party apps (Section~\ref{sec:other_eco}). Finally, we examine the limitation of our work~(Section~\ref{sec:limitation}).

\subsection{Broader Impact}\label{sec:broad}

\paragraph{Distributed Denial-of-Service (DDoS) attack}
When APIs provided to \oai by third-party are leaked, attackers can gain detailed knowledge about the API's structure, request types, and data processing methods through reconnaissance. 
This allows them to precisely design their attacks, targeting resource-intensive operations or vulnerable endpoints. Moreover, by repeatedly invoking API functions that consume a lot of computational resources, attackers may quickly deplete the server's processing capabilities, resulting in a Distributed Denial-of-Service (DDoS) attack~\cite{jaafar2019review}. 
Once succeeded, it could lead to the collapse of the third-party server, consequently preventing users from using corresponding plugins or apps in \gpt.

\paragraph{API misuse}
Attackers can exploit leaked APIs for illicit financial gains~\cite{amann2018systematic,sven2019investigating}. For example, integrating these leaked APIs into their own apps without authorization from the third-party service provider.
This not only constitutes copyright infringement but also violates business regulations as it disrupts the fair competitive market environment.

\paragraph{Fake or malicious plugins}
Previous work~\cite{iqbal2023llm} has confirmed that when plugin manifests are made public, attackers can exploit the meta-data of a plugin to create an identical one and upload it to the GPT plugin store. Unaware users might choose this fake malicious plugin, leading to the leakage of sensitive information. This type of attack leverages the appearance and functionality of legitimate plugins, enticing users to download and use them, thus achieving the attackers' objectives.

\subsection{Legacy Plugins in GPT Store}\label{sec:gptstore}

\begin{figure}[t]
    \centering
    \includegraphics[width=0.48\textwidth,trim={15cm 4cm 15cm 3cm},clip]{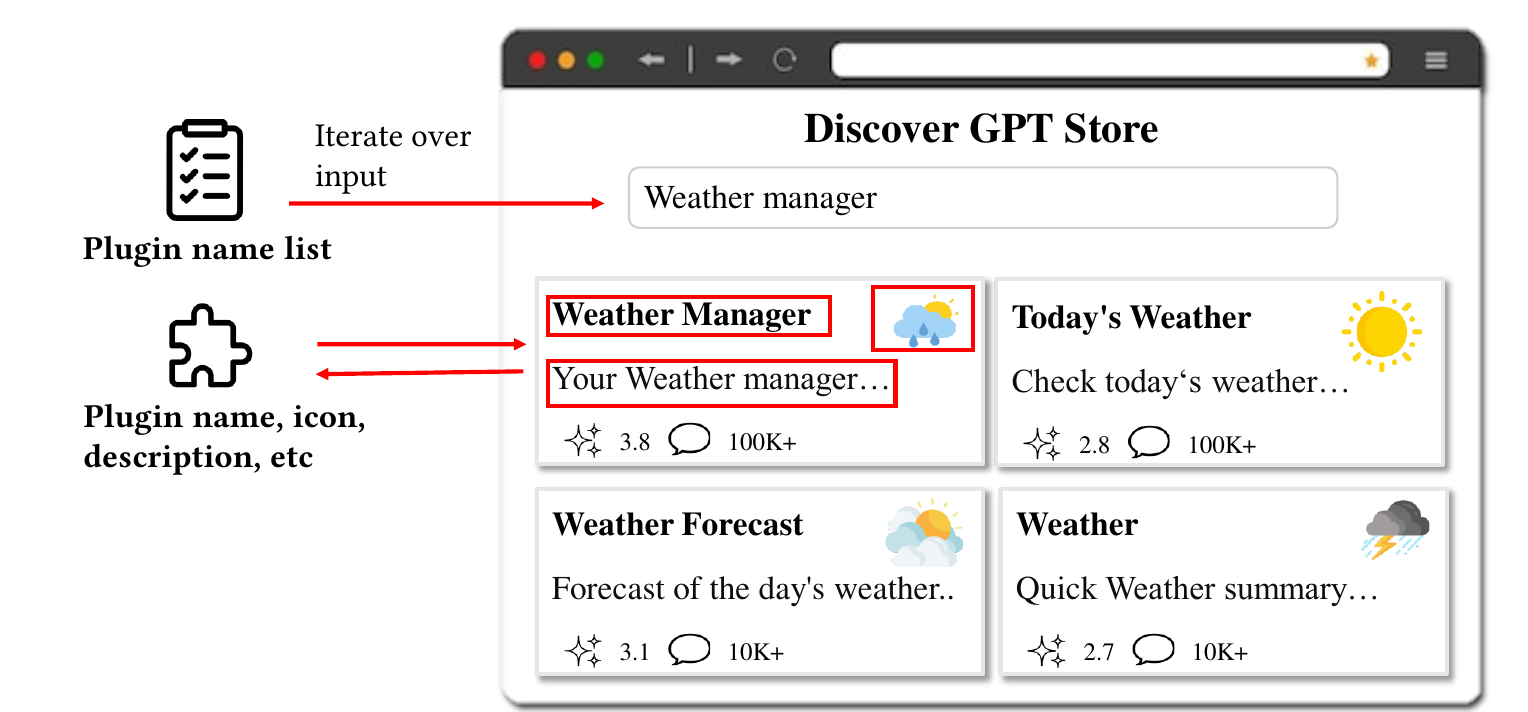}
    \vspace{-0.5cm}
    \caption{The process of detecting plugins in the GPT store}
    \vspace{-0.35cm}
    \label{fig:hunter_ui}
\end{figure}

Emerging as an evolution of \gpt plugin store, the GPT store empowers developers to release their personally developed extensions. Diverging from the plugin store, the GPT store not only facilitates the incorporation of third-party APIs but also strongly promotes the development of extensions through prompts. This innovative strategy notably diminishes the GPT store's reliance on external APIs.

However, after conducting preliminary screening, we discover that plugins have not disappeared following the close of the \gpt plugin store. Instead, some plugins have transformed into GPTs and continue to exist within the GPT store. To investigate the survival of plugins in this new store, we design an automated detection process based on Selenium~\cite{selenium2024dev}, a web testing tool.
Due to the implementation of anti-scraping technologies by the GPT store, we are unable to conduct automated testing directly on it. Hence, we employ GPTs Hunter~\cite{gpthunter23Discover}, an integrated dataset platform encompassing all existing GPTs.
We first locate the input box of GPTs Hunter and enter the name of the target plugin. GPTs Hunter then returns several GPTs that are most relevant to that name as shown in Figure~\ref{fig:hunter_ui}. We then locate and retrieve the icons, descriptions, and names of these GPTs. Next, we compare the icons of the target plugin with those of GPTs using Perceptual Hash Algorithm~\cite{weng2011secure} to assess similarity. The algorithm efficiently identifies and compares image content by generating hash values that are robust to visually similar images.

\paragraph{Evaluation}
To verify whether the GPTs found in the GPT store are identical to the original plugins, we employ a reverse verification method. Among the plugins that could obtain valid information through external API calls, we find that 70 of them have corresponding GPTs. We compare results by inputting identical prompts into both the plugin APIs and GPTs. Aside from minor stylistic differences in language embellishment within ChatGPT, the results are essentially identical. This further confirms that these GPTs and the previous plugins indeed utilize the same API.

Finally, we discover that out of 1038 plugins, 417 are still available in the GPT store as GPTs. Among them, 70 have previously leaked manifest files, and 41 are still externally accessible through external API requests.

\subsection{Recommendations} \label{sec:recommendations}
We propose our recommendations to various parties in the GPT ecosystem who may be affected by security exposures.
\paragraph{OpenAI} As a platform, there is a responsibility and obligation to maintain the security of the ecosystem. We propose the following recommendations for \oai.

\emph{Enhanced third-party API security.}
A comprehensive security assessment of all third-party APIs should be conducted. This includes auditing the authentication mechanisms, data encryption methods, access control policies, log and monitoring practices of the APIs. Access control to API endpoints should be strengthened to ensure that only authorized users or systems can access sensitive data or perform critical operations.

\emph{Review of data submitted by third-party services.}
\gpt should ensure the data submitted by third-party services is accurate and reliable by verifying the sources of the data, checking for discrepancies or anomalies, and validating the integrity of the received data. Additionally, third-party services should be assessed regarding compliance with relevant laws, regulations, and standards, especially those concerning data protection and privacy.

\emph{Provide security guidance and support.}
The store operator should provide security best practices and guidance for developers and users utilizing third-party services. This includes how to integrate and use third-party services within GPT safely, as well as how to report and respond to security issues when they arise.

\emph{Strengthening vulnerability management.}
The store operator should establish a systematic vulnerability management program to identify, assess, and remediate vulnerabilities that may affect the security of integrated third-party services. They should ensure rapid and effective mitigation of any security issues caused by third-party services.

\paragraph{Third-party developers}
Third-party developers should also enhance the security of their services to ensure seamless integration with other platforms and systems.

\emph{API authentication.}
Developers should protect services from unauthorized access by implementing robust authentication and controls on all API endpoints, such as using OAuth, JWT, or other secure token services.

\emph{Manage files effectively under the domain.}
Developers should manage files under their domains to prevent data leaks, especially files that store critical configuration information. There should be mechanisms enforced to define different access permissions for different roles. For example, plugin configuration files should be accessible only by the system, and not by regular users.

\emph{Regularly updated information.}
When the content of a plugin changes, such as alterations to the legal documentation URL, developers should promptly report the change to the platform. Additionally, developers should avoid engaging in unfair ranking competition, such as adding ``A'' to the beginning of a name to achieve a higher placement, as this can mislead users.

\paragraph{Software engineering researcher}
Our work would also encourage software engineering researchers to further explore this new architecture of LLM-centered third-party apps.

\emph{Security threat modeling.}
Researchers could examine the usage scenarios of third-party applications within LLMs and identify potential sources of security threats. This involves distinguishing sensitive data and corresponding access permissions. Additionally, it is crucial to study how to manage both input and output data, as well as to identify potential attack surfaces.

\emph{Privacy compliance assessment.}
The compliance assessment is another topic for researchers. 
Auditing the documentation provided by third-party applications can unveil whether their deployment adheres to data protection regulations such as GDPR and CCPA. For example, they can examine data access permissions, data processing, and data storage methods to ensure that all operations are conducted within the framework of the regulations, thereby avoiding the risks of data misuse or leakage. Previous studies~\cite{yan2024investigating,xie2022scrutinizing} that have identified inconsistencies in fields such as Android and virtual personal assistants can be adapted for LLM-centered apps.

\subsection{Other LLM Ecosystems}\label{sec:other_eco}
At the moment of this work being conducted, \gpt is the most popular LLM platform for integrating third-party applications, having the largest user base and the most active developer community. 
For that reason, our study focuses on \gpt-based app.
Nevertheless, our assessment can still be generalized to other LLM app ecosystems that adopt a similar integration model of leveraging APIs to integrate third-party applications. 
Below we provide three examples of existing LLM ecosystems that can be comprehensively assessed by extending our work. 

\paragraph{Coze} Coze~\cite{coze} is a development platform launched by ByteDance that supports the integration and use of multiple LLMs. It allows developers to choose different LLMs based on their specific needs to power their chatbots or applications. Similar to OpenAI's plugin system, Coze offers a centralized platform where users can publish and download chatbots and plugins created by other developers, thereby extending their functionalities and utilizing services provided by third-party APIs~\cite{cozeai}.

\paragraph{Gemini} Google has integrated its Gemini~\cite{gemini} LLM into its broader ecosystem, serving as the backbone for various apps available through platforms like Google Workspace and Android. 
It does not function as a traditional app store, but provides third-party developers with the tools to create apps that harness the power of Gemini's capabilities.

\paragraph{Poe} Poe~\cite{poe} is an AI chat platform launched by Quora, featuring a plugin marketplace where developers can publish the plugins or extensions they have created. These plugins typically communicate with third-party servers via RESTful APIs, enabling them to add specific functionalities or knowledge based on the AI model on the Poe platform, catering to the diverse needs of its users.

\subsection{Limitation}\label{sec:limitation}
To the best of our knowledge, this work is the first comprehensive security measurement in third-party services within the \gpt app ecosystem. However, the current work carries several limitations that should be addressed in future work.

First, since the debut of the \gpt plugin store, it only has an 11-month lifespan when OpenAI phased it out. Our monitoring covers only the last four months of the \gpt plugin store because the access to \gpt plugins was exclusive to selected users at the beginning.
For that reason, our analysis of the plugin characteristics may be limited when compared with large-scale measurements in other ecosystems such as Android and iOS.   

Second, it is worth noting that our dataset may not encompass every single plugin listed on the \gpt plugin store at any given time. This limitation arises from the possible plugins with extremely short lifespans, i.e., plugins released on \gpt plugin store and quickly de-listed within our data collection interval. 

Additionally, we have reported security issues such as file leaks and unauthenticated API access exclusively to \oai. We have not communicated these to each third-party developer due to the large number, limiting our insights into the duration of the vulnerability resolution process. 
In the future, we plan to conduct more comprehensive security analyses of the newly introduced \gpt store and communicate with both the system and third-party developers towards a secure \gpt ecosystem.

\section{related work}
There are numerous work conducting large-scale measurement and analysis for the security of browser extensions or app stores~\cite{haupert2018honey, papageorgiou2018security,fok2022large,obaidat2020xss,verderame2020appregator,chen2018mystique}. Some recent research has been dedicated to studying the security of LLMs from multiple perspectives~\cite{su2024gpt,zhang2024first,ullah2023can}. To the best of our knowledge, our work is the first and largest exhaustive analysis made for the plugin store of the LLM application ecosystem.

\paragraph{Large-scale extensions or app store monitoring analysis}
Wang et al.~\cite{wang2022characterizing} conduct the first systematic study on cryptocurrency-themed malicious browser extensions, monitoring and analyzing 3,600 extensions to identify 186 malicious ones. They reveal their distribution channels, life cycles, developers, and illicit behaviors, shedding light on the characteristics of these extensions.
Reitinger et al.~\cite{reitinger2023analysis} provide an in-depth analysis of the evolution of Google Ads settings, demonstrating their progression towards being more updated, personalized, precise, and selectively filtered to improve user privacy and experience.
Wang et al.~\cite{wang2018beyond} present a large-scale comparative study of Chinese Android app markets, going beyond Google Play to highlight the unique characteristics and dynamics within China's mobile app ecosystem.

\paragraph{Security risk of application}
Numerous studies~\cite{wan2024safe,zhou2022uncovering,Wang2023understanding} have concentrated on identifying security vulnerabilities within applications. For broken access control vulnerability, Parkinson et al.~\cite{Parkinson2022A} develop methodologies for identifying and mitigating potential permission issues through empirical security analysis of various access control systems to prevent data breaches and unintended modifications.
For Identification and Authentication Failures vulnerability, Wang et al.~\cite{Wang2023understanding} through a systematic analysis of the reasons behind the failure of security proofs in multi-factor authentication schemes for mobile devices, has uncovered the challenges in designing secure and efficient multi-factor authentication systems.

\paragraph{Security analysis of LLMs}
Yao et al.~\cite{yao2024survey} delve into the dual-edged nature of LLMs, highlighting their revolutionary applications for security enhancement. The author demonstrates LLM exploitation by human-like reasoning and discusses the urgent need for further research towards a more robust defense mechanism.
Kshetri et al.~\cite{kshetri2023cybercrime} explore how LLMs pose significant threats to security and privacy. This work underscores the need for comprehensive safety and privacy measures in training and deploying these models.

Unlike previous studies, our research focuses on the \gpt plugin ecosystem and conducts a comprehensive analysis to assess plugin security. We design a three-layer security assessment model to evaluate the security exposure of plugins. Additionally, we uncover API authentication vulnerabilities in both the \gpt plugin store and GPT store, offering guidance for developers and store operators to enhance security within the LLM application ecosystem.

\section{conclusion}
In this work, we have conducted the first comprehensive study of the \gpt app ecosystem from the perspectives of distribution, deployment, and security. 
We characterize all existing plugins, revealing their functionality distribution, which can serve as a reference for developers in deciding their future endeavors, and for store operators in offering personalized services.
We also investigate the deployment and execution models of the plugins through reverse engineering. 
Based on these models, we identify five types of potential exposures and propose a three-layer security assessment to explore the security landscape of the \gpt app ecosystem. 
Our findings would encourage both \oai and third-party developers to collaboratively maintain and develop a healthy LLM app ecosystem.

\paragraph{Availability} The source code of our work and relevant artifacts are available online~\cite{github}.
\section*{Acknowledgement}

We thank anonymous reviewers for their insightful comments. 
This research has been partially supported by Australian Research Council Discovery Projects~(DP230101196, DP240103068).

\bibliographystyle{ACM-Reference-Format}
\bibliography{sample-base}


\end{document}